# Ongoing EEG artifact correction using blind source separation

Nicole Ille[a,*], Yoshiaki Nakao[b], Yano Shumpei[b], Toshiyuki Taura[b], Arndt Ebert[a], Harald Bornfleth[a], Suguru Asagi[c], Kanoko Kozawa[c], Izumi Itabashi[c], Takafumi Sato[c], Rie Sakuraba[c], Rie Tsuda[c], Yosuke Kakisaka[d], Kazutaka Jin[d], Nobukazu Nakasato[d]

[a]BESA GmbH, Gräfelfing, Germany
[b]Nihon Kohden Corporation, Tokyo, Japan
[c]Clinical Physiological Center, Tohoku University Hospital, Sendai, Japan
[d]Department of Epileptology, Tohoku University Graduate School of Medicine, Sendai, Japan

*Corresponding author: Nicole Ille, Besa GmbH, Gräfelfing, Germany. E-mail: nille@besa.de

## Abstract

**Objective:** Analysis of the electroencephalogram (EEG) for epileptic spike and seizure detection or brain-computer interfaces can be severely hampered by the presence of artifacts. The aim of this study is to describe and evaluate a fast automatic algorithm for ongoing correction of artifacts in continuous EEG recordings, which can be applied offline and online.

**Methods:** The automatic algorithm for ongoing correction of artifacts is based on fast blind source separation. It uses a sliding window technique with overlapping epochs and features in the spatial, temporal and frequency domain to detect and correct ocular, cardiac, muscle and powerline artifacts.

**Results:** The approach was validated in an independent evaluation study on publicly available continuous EEG data with 2035 marked artifacts. Validation confirmed that 88% of the artifacts could be removed successfully (ocular: 81%, cardiac: 84%, muscle: 98%, powerline: 100%). It outperformed state-of-the-art algorithms both in terms of artifact reduction rates and computation time.

**Conclusions:** Fast ongoing artifact correction successfully removed a good proportion of artifacts, while preserving most of the EEG signals.

**Significance:** The presented algorithm may be useful for ongoing correction of artifacts, e.g., in online systems for epileptic spike and seizure detection or brain-computer interfaces.

**Keywords**: online artifact removal, electroencephalogram, blind source separation, independent component analysis, brain-computer interface, epileptic spike and seizure detection

# 1. Introduction

Analysis of continuous and event-related recordings of the electroencephalogram (EEG) can be severely hampered by the presence of physiological artifacts such as blinks, eye movement, cardiac or muscle activity and non-physiological artifacts caused for example by 50 Hz or 60 Hz powerline interference.

Many approaches have been described over the past years to reduce artifact contamination while trying to preserve most of the brain activity even if this is correlated with artifact activity. A large amount of published methods is based on blind source separation (BSS) or independent component analysis (ICA) since the introduction of ICA for ocular artifact correction (Vigário, 1997; Jung et al., 1998). Other successful approaches use for example spatial filters modelling artifact and brain activity (Berg and Scherg, 1991, 1994; Ille et al., 1997, 2002), spatially constrained ICA (SCICA) (Ille, 2001; Ille et al., 2001; Hesse and James, 2006), or hybrid approaches like BSS/ICA in combination with wavelet transformation (WT) (Castellanos and Makarov, 2006; Mammone et al., 2012; Mammone and Morabito, 2014; Akhtar et al., 2012). For comprehensive reviews of EEG artifact removal methods, see for example Kaya (2022), Urigüen and Garcia-Zapirain (2015), Islam et al. (2016).

Spatial filters and SCICA rely on the existence of an a priori defined artifact model, in general a topography, restricting the method to the correction of spatially fixed artifacts like ocular or cardiac artifacts (Ille et al., 2002). BSS/ICA approaches, on the other hand, can be applied to a wider range of artifact types including muscle and non-physiological artifacts. To this end, an EEG segment is transformed to BSS/ICA components using a suitable algorithm. Artifact components are either identified manually, semi-automated or automatized using method-specific heuristics. Finally, the clean EEG segment is reconstructed without the identified artifact components.

There are multiple automatized BSS-based artifact correction pipelines, e.g., FASTER (Nolan et al., 2010), ADJUST (Mognon et al., 2011) and MARA (Winkler et al., 2011, 2014) for ERP pre-processing or a system for automatic artifact removal in ictal EEG (LeVan et al., 2006). As ICA is calculated for longer portions of the raw data or even for the whole data set, these approaches are only suitable for offline processing. Online processing, on the other hand, is required for brain-computer interfaces (BCI), online spike and seizure detection or patient monitoring in critical care units.

Only few BSS-based artifact correction methods have been adapted for online usage, e.g. Halder et al. (2007) and FORCe (Daly et al., 2015) in the context of BCI. These use in general short overlapping sliding windows of 1 to 3 s length. If the sliding window is too small, there may not be enough samples to reliably estimate the weights of the BSS unmixing matrix (Delorme and Makeig, 2004; Onton et al., 2006; Korats et al., 2012). Therefore, the OCARTA method (Breuer et al., 2014) uses a longer sliding window with variable overlap, which is advanced to new data whenever BSS calculation is finished. None of these approaches, however, consider boundaries of corrected segments. Only in the methods reported in Wallstrom et al. (2004) and Nierenberg et al. (2015), the corrected segments are aligned using a weighted averaging technique to avoid discontinuities at segment boundaries in the corrected data.

In this paper, we present an online-capable BSS-based algorithm for correction of ocular, cardiac, muscle and powerline artifacts in continuous EEG, which uses a longer sliding window to reliably estimate the weights of the BSS unmixing matrix. The sliding window has a fixed overlap and covers 75% of past data. Corrected segments are smoothly aligned by means of the weighted averaging method described in Wallstrom et al. (2004) in order to create a corrected EEG without discontinuities for further review or processing. To identify BSS artifact components, intuitive heuristic criteria based on features in the temporal, spatial and frequency domain are used. No user interaction and no additional recorded reference channels are required.

The algorithm has been independently validated by expert neurologists using previously unseen validation data from the publicly available TUH EEG Artifact Corpus (Hamid et al., 2020).

## 2. Methods

Prior to ICA, recorded EEG data is filtered with a time constant of 0.3 s and a high-cutoff filter of 70 Hz. Then a principal component analysis (PCA) with dimension reduction is applied retaining all PCA components explaining $\geq 1\%$ of the variance in the data.

Independent components are calculated on the pre-processed data using a fast orthogonal extended infomax algorithm (OgExtInf) (Ille, 2023) combining classical extended infomax (Lee et al., 1999) with a fast fully-multiplicative orthogonal-group based update scheme of the weight matrix (Fiori, 2003). Like the classical extended infomax, the OgExtInf method is suitable for separating cardiac, ocular and powerline artifact components.

Ocular artifact components are detected in the OgExtInf waveforms using synthetic electrooculogram (EOG) channels sensitive for blink, horizontal and vertical eye movement. Synthetic EOG channels are calculated from the recorded EEG data by means of a spatial filter taking into account activity in multiple brain regions (Scherg et al., 2002) without the need of recording additional horizontal and vertical EOG channels. EOG artifact components are detected by both temporal and spatial correlation with the synthetic EOG components as opposed to previous artifact reduction methods that exploited correlation either with recorded EOG waveforms (Wallstrom et al., 2004; Joyce et al., 2004; Nolan et al., 2010; Chaumon et al., 2015) or topographies (Li et al., 2006; Campos Viola et al., 2009).

Cardiac artifact components can be detected with or without reference to a recorded electrocardiogram (ECG) channel. If an ECG channel is available, cardiac artifact components are detected by correlation between energy transformed recorded ECG waveform and energy transformed ICA waveform (De Vos et al., 2011). If no ECG channel is available, cardiac artifact components are identified by frequency and periodicity of autocorrelation of the energy transformed ICA waveform, and the location of the corresponding equivalent current dipole. In previous studies, for example, the frequency of power spectral density (PSD) (Tamburro et al., 2018) and periodicity measures derived from the time course (Tamburro et al., 2018) or the continuous wavelet transform (CWT) of the ICA waveform (Hamaneh et al., 2014) were used for cardiac artifact detection.

50 Hz or 60 Hz powerline artifact components are also detected by frequency. Powerline artifacts are corrected only in affected EEG channels.

Before detection of muscle artifacts, ocular, cardiac and powerline EEG signals are reconstructed from the identified artifact components and removed from the data. The resulting data is high-pass filtered (> 10 Hz) to focus on muscle activity (Frølich and Dowding, 2018). Finally, the Second Order Blind Identification (SOBI) algorithm (Belouchrani et al., 1997), which is a widely used BSS algorithm, is calculated, and muscle artifacts are detected in the SOBI components by low autocorrelation (Wim De Clercq et al., 2006; Vos et al., 2010; Chaumon et al., 2015). The whole sequence of artifact correction is summarized in Fig. 1.

For ongoing correction, data is processed repeatedly in blocks of 2 s. Internally, 4 data blocks of length 2 s are accumulated: the current and past three blocks. Whenever a new block of 2 s is added, the oldest 2 s block is discarded. Artifacts are detected in the 8 s data buffer. In each step, the corrected $3^{rd}$ buffer is appended to the continuous, corrected EEG resulting in a delay of 2 s in ongoing correction mode. To avoid discontinuities at buffer boundaries corrected data is combined with the overlapping corrected signal of the previous step by weighted averaging according to Wallstrom et al. (2004). The process of ongoing artifact correction is illustrated in Fig. 2. The presented algorithm is implemented in the BESA Artifact Module (BAM) of BESA GmbH (Gräfelfing, Germany).

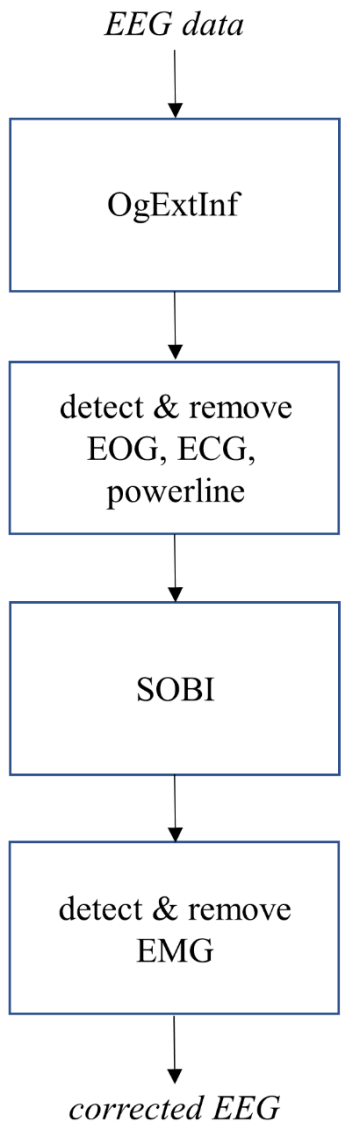

Figure 1: Sequence of artifact correction.

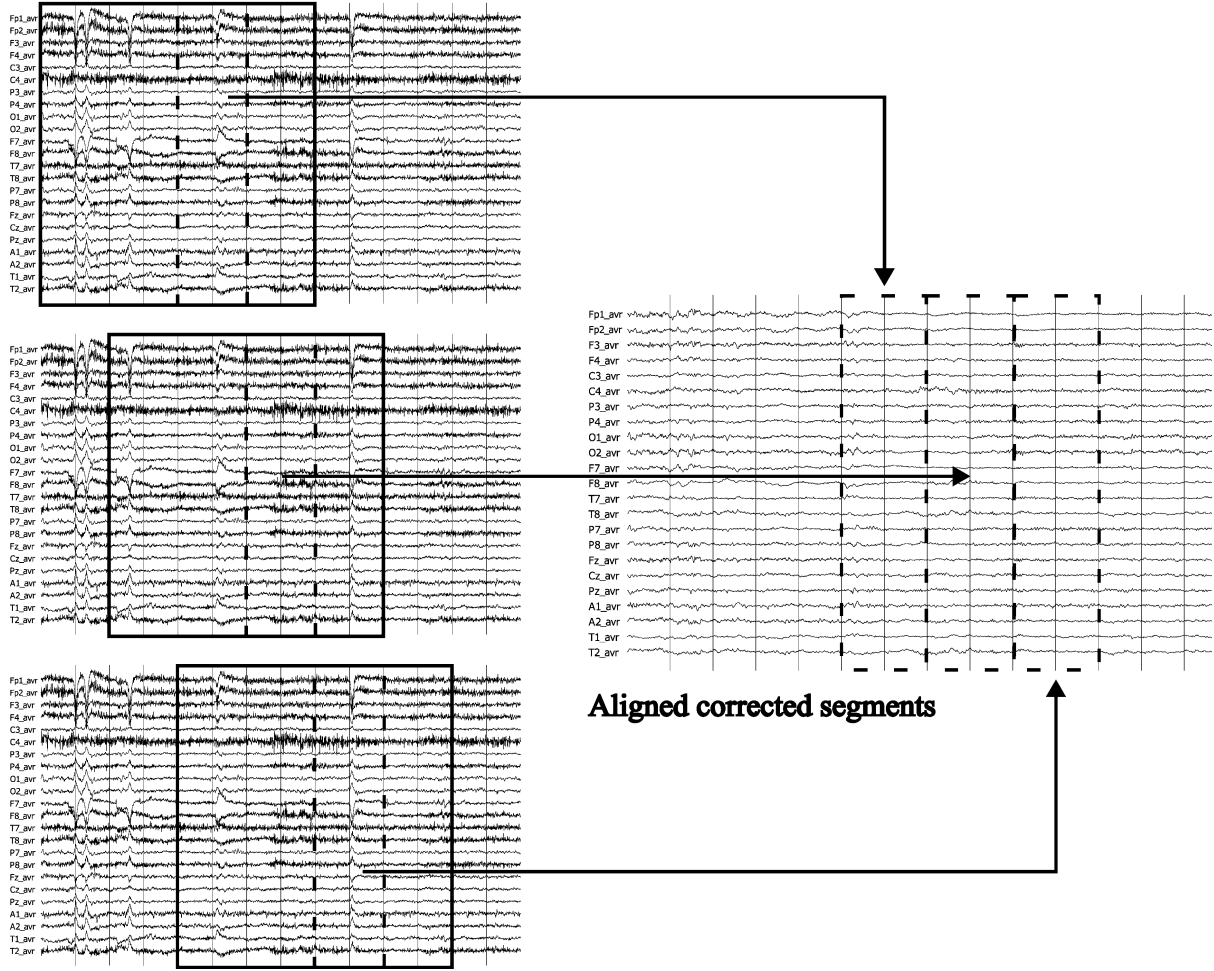

Figure 2: Process of ongoing artifact correction: Artifacts are detected and corrected in the 8 s framed epochs. The 2 s segments within the dashed frames are smoothly aligned to the continuous, corrected EEG.

## 3. Results

Ongoing artifact correction with the BAM algorithm was validated in the Department of Epileptology, Tohoku University Hospital, Sendai, Japan, using the BESA Artifact Module plugin of the Nihon Kohden EEG Review software (QP-112A/AK). For validation, 22 previously unseen off-line EEG recordings of approximate length 10 minutes were taken from the open-source TUH EEG Artifact Corpus (Hamid et al., 2020). The 22 EEGs consist of 22 to 25 recording channels (mean=23) according to the 10-20 system. Sampling rates range from 256 Hz to 1024 Hz.

Before analysis, the TUH EEG data files were converted from European Data Format (EDF+) to Nihon Kohden data format using the BESA Converter Module of the Nihon Kohden EEG software (EEG-1260A/EEG-1200K/QP-112AK). During conversion data was resampled to fixed sampling rates, resulting in 19 EEGs with 500 Hz and 3 EEGs with 1000 Hz sampling rate.

The 22 EEGs were annotated with a total of 2035 artifacts including 977 ocular, 551 muscle, 297 cardiac and 210 powerline artifacts. In addition, data contains normal EEG background, seizures and spikes. Powerline frequency is always 60 Hz in this data set. The 22 EEGs were corrected for artifacts using the ongoing BAM algorithm. No recorded ECG channel was used for ECG correction in this validation study. In Figs. 3 and 4, examples of original and corrected segments during seizures are shown.

The quality of ongoing artifact reduction with the BAM algorithm was evaluated visually by 6 trained electroencephalographers. 5 of the expert reviewers had more than 3 years and one had 1 year of experience of analyzing EEG data. For evaluation, the data sets were split evenly between the reviewers. The reviewers compared the original and the corrected EEG signals and judged annotation by annotation whether an artifact was completely or mostly removed (good correction) or not. The reviewers' criterion for a good correction was a major improvement of the signal quality corresponding to a visually estimated signal-to-noise ratio greater than 1 after correction. In addition, the reviewers examined separately in the whole recording whether EEG signals were distorted by artifact correction.

In Table 1, the reduction rate, i.e., the percentage of good corrections, is reported for each artifact category and in total. Overall, 88% of all annotated artifacts were eliminated successfully from the EEG signals, while 12% of artifacts were only partially removed including few cases where artifacts are not at all reduced.

In addition, 62 distortions of EEG signals were observed in the 220 minutes of EEG data (see Table 2). This corresponds to a frequency of distortion of 16.9 per hour. Distortions were mainly related to EOG correction (58 of 62). No distortion was reported for ECG and powerline correction. Observed distortions included, for example, alpha rhythm introduced in frontal channels by EOG correction and slow or background waves that were removed as EOG.

To make sure that the proposed algorithm is fast enough for online processing, the processing time was evaluated for 5 randomly selected EEG data sets of the evaluation study with 23 recording channels and 500 Hz sampling rate. This can be considered standard clinical EEG recording conditions. Computation time for one data block of 8 s was found to be on average 0.135 s ($\pm$ 0.027 s) on an Intel Core i7-7700 CPU @ 3.6 GHz with 4 cores and 16 GB RAM (min: 0.086 s, max: 0.331 s, median: 0.128 s). Thus, ongoing artifact reduction with the BAM algorithm is considerably faster than the 2 s needed to accumulate new data and introduces only a negligible additional delay.

| Artifact type | Nr. of artifacts | Reduction Rate (%) |
| --- | --- | --- |
| EOG | 977 | 81 |
| EMG | 551 | 98 |
| ECG | 297 | 84 |
| Powerline | 210 | 100 |
| Total | 2035 | 88 |

Table 1: Number of annotated artifacts and reduction rate (percentage of good corrections) for each artifact category and in total.

| Artifact type | Nr. of distortions | Frequency (per h) |
| --- | --- | --- |
| EOG | 58 | 15.8 |
| EMG | 4 | 1.1 |
| ECG | 0 | 0 |
| Powerline | 0 | 0 |
| Total | 62 | 16.9 |

Table 2: Absolute number of distortions of EEG signals introduced by artifact correction for each artifact category and in total.

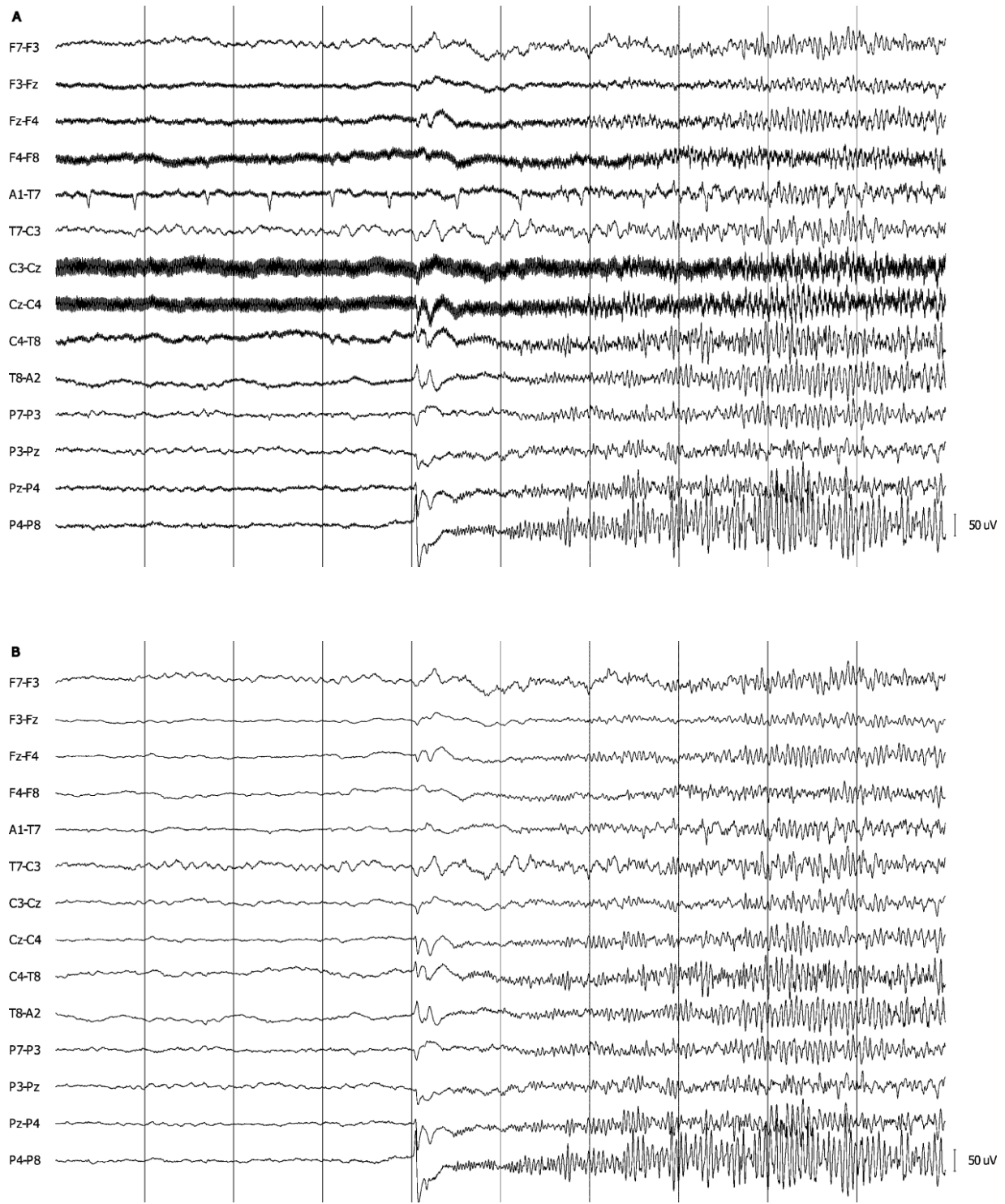

Figure 3: (A) A 10 s segment at the beginning of a seizure affected by cardiac and powerline artifacts. (B) The ongoing BAM algorithm removed the artifacts, and the onset of the seizure is clearly visible.

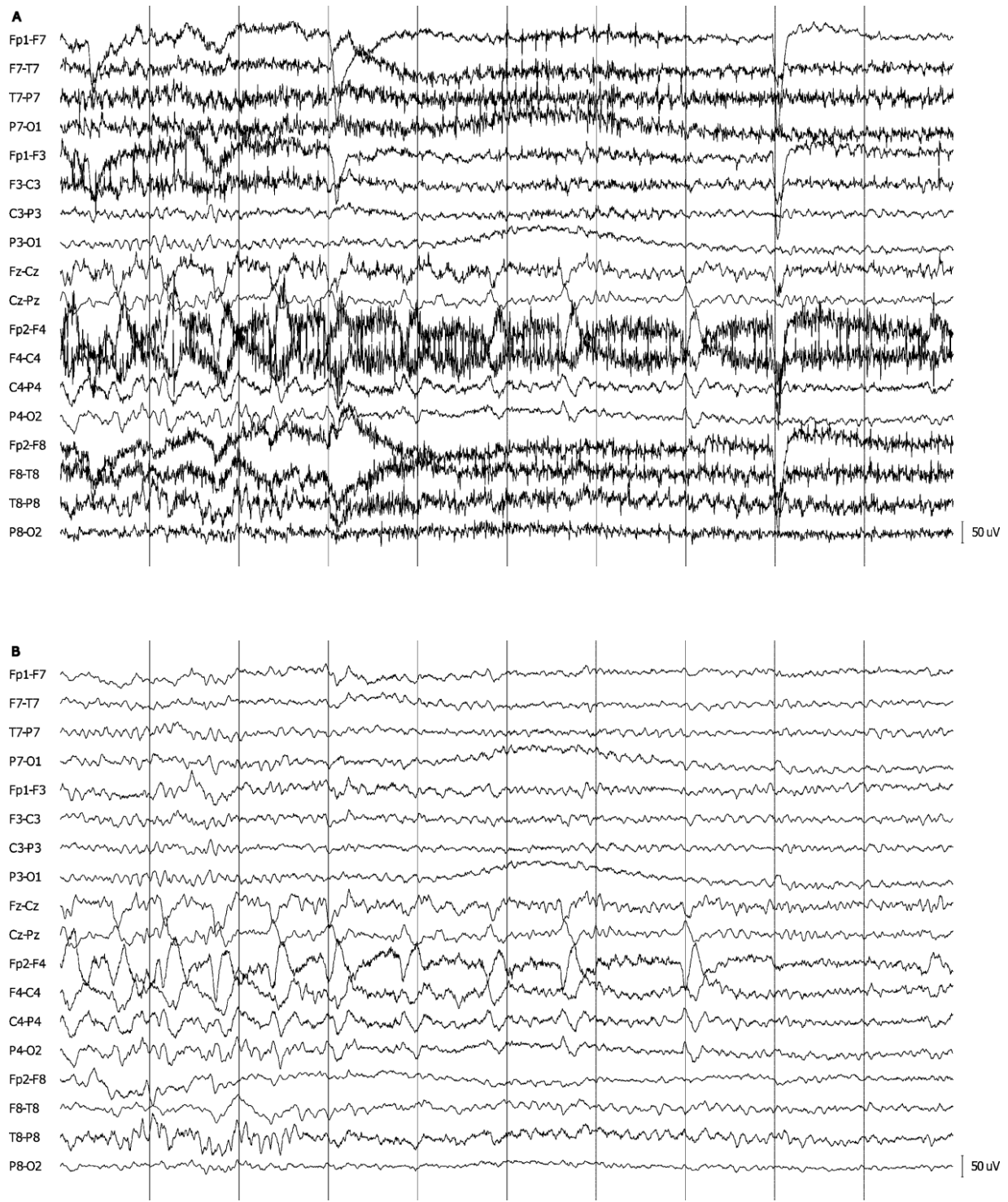

Figure 4: (A) A 10 s segment with a seizure heavily contaminated by ocular and muscle artifacts. (B) After processing, the artifacts are corrected, and the end of the seizure is discernible.

# 4. Discussion

## 4.1. Ongoing versus single data segment

In this work, ongoing artifact correction with the BAM algorithm was introduced. The presented method can also be applied to the correction of single data segments. However, only the sliding window approach presented here and the overlap of epochs allow to combine the corrected epochs into a single and unique continuous, corrected realization of the data.

## 4.2. Quality of artifact correction

In our evaluation, it was found that 88% of the artifacts in the EEG signals can be removed successfully by ongoing BAM (ocular: 81%, cardiac: 84%, muscle: 98%, powerline: 100%). This is considerably better than previously reported artifact reduction rates in EEG signals using BSS-based artifact correction.

In a previous study (LeVan et al., 2006), visual inspection of an expert neurologist using similar criteria showed that only 48.8% of artifacts in 205 seizure segments with a duration of 30 s each were removed mostly or with a major improvement of the EEG signals. In the same study, major or minor attenuation of EEG activity by artifact correction was found in 15.5% of the seizure segments. This corresponds to a frequency of 18.6 distorted segments per hour which is in line with the 16.9 single distortions per hour observed in this evaluation.

For the FORCe method, visual inspection of two trained reviewers revealed that 62% of the muscle artifacts and a statistically significant amount of 75% of the blinks in the original EEG could be removed (Daly et al., 2015). In the same study, a comparison with other state-of-the-art BSS-based methods for BCI artifact handling, LAMIC (Nicolaou and Nasuto, 2007) and FASTER (Nolan et al., 2010), was performed (see Table 3). Here LAMIC removed only 40% of the blinks and 3% of muscle artifacts. FASTER removed only 9% of the blinks and 48% of muscle artifacts in the original EEG, although more than 94% sensitivity is reported for this method for the detection of eye and muscle artifacts in simulated data (Nolan et al., 2010).

Many BSS-based studies report classification rates of independent components (as agreement with human experts that labelled the artifact components), instead of artifact reduction rates in EEG signals. Classification rates of independent components are in general larger than 85% for MARA (Winkler et al., 2011, 2014) up to 95% for Adjust (Mognon et al., 2011; Radüntz et al., 2017) and even 99% for ocular artifacts (Halder et al., 2007). However, classification rates of independent components cannot be compared to reduction rates of artifacts in EEG signals as they neither indicate completeness of artifact correction nor possible distortions of EEG signals by artifact removal.

In this study, ocular artifact correction has the lowest observed reduction rate (81%) and also attributes to most of the EEG signal distortions (94%). This is mainly due to characteristics of ICA. On the one hand, ICA may separate artifact activity, especially ocular activity, into multiple components, which can be hard to detect for automatic algorithms as well as human raters (Wallstrom et al., 2004). On the other hand, components may contain a mixture of artifact and non-artifact activity, e.g., alpha rhythm in a blink, which may either lead to distortion in the corrected signal or classification of the component as non-artifactual (LeVan et al., 2006). In order to compensate for this shortcoming of ICA, hybrid approaches with topography-based methods (e.g. Ille et al., 2002) might be beneficial.

|       | LAMIC | FASTER | FORCe |
|-------|-------|--------|-------|
| Blink | 40%   | 9%     | 75%   |
| EMG   | 3%    | 48%    | 62%   |

Table 3: Reduction rates of blinks and muscle artifacts in EEG signals for BSS-based methods LAMIC, FASTER and FORCe. Reduction rates are calculated from mean percentage artifact contaminations reported in Table VI of Daly et al. (2015).

### 4.3. Online processing

Artifact correction with the BAM algorithm has a mean runtime of 0.135 s (± 0.027 s) for data segments of 23 channels and 8 s at a sampling rate of 500 Hz on a standard computer. The FORCe method has a mean runtime of 0.382 s (± 0.076 s) and LAMIC has a mean runtime of 0.226 s (± 0.023 s) for data segments of 16 channels, 1 s and 512 Hz sampling rate on a standard computer (Daly et al., 2015).

In general, ICA calculation is the bottleneck of BSS-based artifact correction algorithms. Although a longer data segment of 8 s is used to improve ICA decomposition, the BAM algorithm is still considerably faster than FORCe or LAMIC, because a fast orthogonal extended infomax algorithm (OgExtInf) (Ille, 2023) is used.

By ongoing artifact correction, an additional delay of 2 s is introduced by data forward buffering. This prevents discontinuities at segment boundaries and creates a unique corrected EEG but limits the use of ongoing BAM for *real-time* online applications, like for example neuroprosthetic control, where a delay of greater than 0.2 s will introduce a noticeable delay and degrade performance (Lauer et al., 2000). If BAM is applied to a single segment under the conditions used in this study, however, it is sufficiently fast to fulfill this real-time condition. Ongoing BAM, on the other hand, may be used for online applications like spike and seizure detection or patient monitoring in critical care units where a delay time of 2 s is acceptable. Moreover, forward buffer time could be further reduced to 0.5 or 1 s.

## 5. Conclusion

This paper presents the BAM algorithm for ongoing correction of artifacts in continuous EEG recordings. The algorithm requires no user interaction and no additional recorded reference channels and generalizes to different EEG studies and electrode setups. The BSS-based algorithm uses intuitive and characteristic features in the temporal, spatial and frequency domain to detect and correct ocular, cardiac, muscle and powerline artifacts. An independent validation study confirms that 88% of these artifacts are removed successfully from the EEG signals. Experimental results show that ongoing correction is fast enough to be applied to online applications like brain-computer interfaces or systems for epileptic spike and seizure detection.